# Research on Intra-Chip Fusion Deployment and Optimization of Embodied Intelligence Business Operator NPU

Yuchen Zhu[1], Longxiang Yin[2],Wanyu Wang[4],Jieke Lin[4],Guoqiang Zou[4],Zirui Cao[3],Yuling Yuan[3],Xiaolan Fan[2],Lifen Chen[2],Hao Zheng[1],Qizhang He[4],Hongyu Zhou[4],Chunhai Yu[4]

*(1. Beijing Information Science and Technology University, Beijing 102206, China; 2. Institute of Computing Technology, Chinese Academy of Sciences, Beijing 100190, China; 3. ShanghaiTech University; Shanghai 201210, China 4. Hangzhou Institute for Advanced Study, UCAS, Hangzhou 310024, China)*

**Abstract:** Embodied intelligent computing systems are characterized by the integration of perception, computation and control. In traditional schemes, perception, computation and control tasks are deployed on separate processors, which causes frequent data migration, high end-to-end latency and low hardware computing efficiency. Such drawbacks fail to satisfy the millisecond-level real-time response demands of embodied intelligent systems in highly dynamic scenarios. Furthermore, most existing operator optimization methods are developed on foreign GPU platforms, and there remains a notable research gap in full-process collaborative optimization for the domestic Phytium-Cambricon heterogeneous architecture.To solve the above problems, this study establishes a domestic heterogeneous computing platform based on Phytium FT-2000/4 main control processor and Cambricon MLU370 intelligent acceleration card, and proposes an NPU on-chip fusion deployment and collaborative optimization strategy for operators throughout perception, computation and control pipelines. Oriented to typical application scenarios of embodied intelligent robots, this paper carries out targeted optimization design: MLU hardware adaptation of motion blur distortion correction operators for high-speed imaging in perception module, lightweight deployment and inference optimization of ViT vision large models in computation module, and customized operator development for inverse kinematics solution of multi-degree-of-freedom motion control in control module. On this basis, a single-card integrated solution adopting on-chip data closed-loop and pipeline collaboration is constructed, enabling the unified execution of full perception-computation-control tasks on a single MLU370 card.Experimental results verify that the proposed scheme reduces the average single-frame latency of the whole process to 18.7 ms with a speedup ratio of 2.89 against NVIDIA Jetson AGX Xavier. The utilization rate of MLU computing units reaches 82.6%, while the processing accuracy is comparable to that of mainstream platforms. This work provides a practical reference for the domestic engineering application of perception-computation-control services in embodied intelligence systems.

## 1 Introduction

With the in-depth integration of artificial intelligence and robotics technologies, embodied intelligence has become a vital development trend for the new generation of unmanned systems, among which perception-computation-control integration serves as its typical technical feature[1-3]. An embodied intelligent system realizes real-time image acquisition and preprocessing in the perception stage, visual feature extraction and scene semantic understanding in the computation stage, as well as motion planning and execution instruction generation in the control stage[4,5]. The service workflow composed of these three stages imposes stringent low-latency requirements on end-to-end processing latency, computational accuracy and system stability[6]. In highly dynamic scenarios such as space-based earth observation and UAV target tracking, excessive latency will directly result in target loss and even trigger safety accidents in severe cases[7]. Hence, low-latency full-process processing capability has become a core performance indicator of embodied intelligent systems.

In traditional unmanned systems, the three subtasks of perception, computation and control are generally deployed separately on different processors including CPU, NPU and FPGA[8]. Such distributed deployment leads to frequent data migration between processors. In some scenarios, data transmission overhead accounts for 30% to 50% of the total process latency, which turns into the major bottleneck restricting system real-time performance[9]. Meanwhile, most existing intelligent computing operator optimization and model deployment technologies are developed based on foreign hardware platforms such as NVIDIA GPU[10]. Their optimization strategies and programming models fail to fully adapt to the architectural characteristics of domestic hardware, thereby limiting the computational efficiency of domestic heterogeneous platforms.

The heterogeneous architecture consisting of Phytium series CPUs and Cambricon MLU series accelerators is one of the most mature and well-established all-domestic computing power solutions in China, possessing broad application prospects in intelligent robots and other fields[11]. Nevertheless, there are still obvious deficiencies in researches on full-process operator collaborative optimization and integrated deployment of perception-computation-control services for this platform, and mature end-to-end optimization solutions are still lacking.

Aiming at the above technical bottlenecks, this study takes typical embodied intelligent robot scenarios as the research object, and conducts research on NPU on-chip fusion deployment and optimization of perception-computation-control service workflow operators based on the domestic heterogeneous platform built by Phytium FT-2000/4 CPU and Cambricon MLU370 NPU. The main research contents are summarized as follows:

(1) In accordance with the service characteristics of the three processing stages, a full-process operator collaborative optimization method is proposed[12]. Algorithm optimization and hardware adaptation for perception, computation and control stages are completed respectively, achieving in-depth matching between operators of each stage and Cambricon MLU370 NPU.

(2) Breaking through the technical limitations of traditional multi-processor distributed deployment, a single NPU on-chip fusion scheme is put forward[13]. By means of operator compatibility adaptation and pipeline collaborative scheduling, closed-loop execution of the entire perception-computation-control workflow is realized within a single MLU370 chip, effectively reducing the proportion of data transmission overhead.

(3) A domestic computing platform supporting embodied intelligent services is established. Multi-dimensional comparative experiments are carried out to verify the performance of the proposed optimization scheme. While guaranteeing processing accuracy, the scheme reduces end-to-end latency and greatly improves hardware computing resource utilization, providing solid technical support for the domestic industrial implementation of embodied intelligent systems[14].

## 2 Related Work

With the rapid development of embodied intelligence and robotics technologies, perception-computation-control integration has evolved into the mainstream development trend of high-real-time unmanned systems[15]. Relevant researches are mainly carried out in four directions: integrated perception-computation-control architecture, NPU operator optimization, lightweight deployment of vision models, and adaptation to domestic heterogeneous platforms.

In terms of perception-computation-control integrated architectures, early studies generally adopted the multi-chip distributed deployment mode based on CPU+NPU, which assigned perception, inference and control tasks to separate hardware for functional decoupling[16]. Nevertheless, such architectures suffer from frequent cross-device data transmission, high end-to-end latency and low resource utilization. In recent years, integrated deployment oriented to low-latency scenarios has become a research hotspot[17]. Scholars have reduced data migration overhead by adopting shared memory, on-chip pipelines and heterogeneous task scheduling strategies[18]. For instance, some studies constructed joint perception-inference pipelines on GPUs to complete image preprocessing and model inference on the same device, which greatly cut down transmission latency[19]. Other researches accelerated control algorithms via FPGA hardware logic and established a GPU-FPGA heterogeneous collaboration framework[20]. However, these solutions still rely on foreign hardware platforms and fail to achieve true closed-loop integration of perception, computation and control within a single NPU, making them incompatible with domestic-oriented, high-real-time and high-integration embodied intelligent application scenarios.

In the field of NPU operator optimization and model deployment, most existing studies are developed based on NVIDIA CUDA, TensorRT, ONNX Runtime and other inference engines[21], focusing on operator fusion, inter-layer optimization, mixed precision computation and dynamic shape adaptation[22]. Intermediate tensor generation can be reduced by means of operator merging, constant folding and memory reuse to boost GPU inference throughput[23]. INT8 quantization can effectively compress model size and lower latency while maintaining acceptable accuracy loss[24]. In recent years, optimization researches targeting domestic NPUs have been on the rise[11]. Platforms including Cambricon and Huawei Ascend have launched self-developed inference frameworks and operator libraries tailored to their own hardware. Even so, most studies only focus on independent acceleration of single vision tasks or individual control algorithms, lacking full-link collaborative optimization schemes covering perception, computation and control, especially end-to-end deployment practices for the Phytium-Cambricon heterogeneous system.

As for lightweight design and edge deployment of large vision models, large vision models such as ViT and Swin Transformer possess prominent accuracy advantages, yet their huge parameter volume and computational complexity hinder their direct deployment on edge devices[25]. Current mainstream lightweight strategies include attention sparsification, embedding dimension compression, model distillation and structural pruning[26]. Specifically, removing redundant attention heads, reducing token dimensions and fusing multi-layer features can drastically reduce computational complexity with limited accuracy degradation[27]. Quantization and operator adaptation further facilitate the deployment of such models on edge NPUs[28]. Nevertheless, existing lightweight researches mostly aim at single vision tasks such as classification and object detection, without establishing logical links with downstream control algorithms, thus failing to fully exploit the overall performance of integrated perception-computation-control system.

In terms of collaborative optimization for domestic heterogeneous platforms, the all-domestic heterogeneous system composed of Phytium CPUs and Cambricon MLUs has been gradually applied in industrial control, intelligent security, robotics and other fields[11]. Current related researches mainly focus on basic aspects including driver adaptation, runtime optimization and basic operator transplantation[29]. For example, CPU-NPU collaboration efficiency is improved by optimizing PCIe bandwidth, adopting asynchronous data transmission and formulating on-chip cache strategies[30], and efficient inference of deep learning models on MLUs has been realized in several studies. There still exist two major deficiencies in existing researches. First, customized operator design and pipeline scheduling for the full perception-computation-control link are insufficient, leading to isolated perception, inference and control tasks and low resource utilization. Second, end-to-end verification of millisecond-level real-time response in highly dynamic scenarios is lacking, which restricts the engineering implementation of embodied intelligent systems.

In summary, existing researches have made certain achievements in single-task acceleration, vision model lightweighting and domestic hardware adaptation, but a complete optimization scheme oriented to domestic heterogeneous platforms that covers the full perception-computation-control link and supports on-chip fusion deployment within a single NPU has not yet been established. To meet the demands for highly dynamic, low-latency and domestically localized embodied intelligence, it is urgent to break through the bottleneck of cross-device data transmission and build an integrated collaborative optimization system for perception, computation and control, so as to provide solid technical support for fully domestic embodied intelligent systems[31].

# 3 Principles of Operator Optimization Algorithms for Full-Process Sensing and Computing Control and Methods for Hardware Adaptation

## 3.1 Optimization of Image Preprocessing Operators in the Perception Module and MLU Hardware Adaptation

Image preprocessing serves as the initial stage of the perception-computation-control workflow. Images captured by embodied intelligent robots in high-speed motion scenarios are prone to motion blur distortion caused by the rolling shutter effect[32]. Without effective correction, this distortion will directly lead to deviations in subsequent visual feature extraction and errors in control commands[33]. Under the traditional CPU serial processing mode, the single-frame processing latency of deep learning-based image distortion correction algorithms exceeds 250 ms, which fails to meet real-time requirements. To address this issue, this study takes the DeepDeblur algorithm as the core and completes the MLU hardware adaptation and inference optimization of the motion blur distortion correction algorithm[34].

### 3.1.1 DeepDeblur Algorithm

DeepDeblur is a deep learning model designed for dynamic blurred image restoration[35]. Its core architecture adopts multi-scale feature fusion and physics-guided Generative Adversarial Network (GAN), which jointly models the blur kernel estimation and clear image generation process[36]. The workflow of the multi-scale feature fusion module is illustrated in Figure 1.

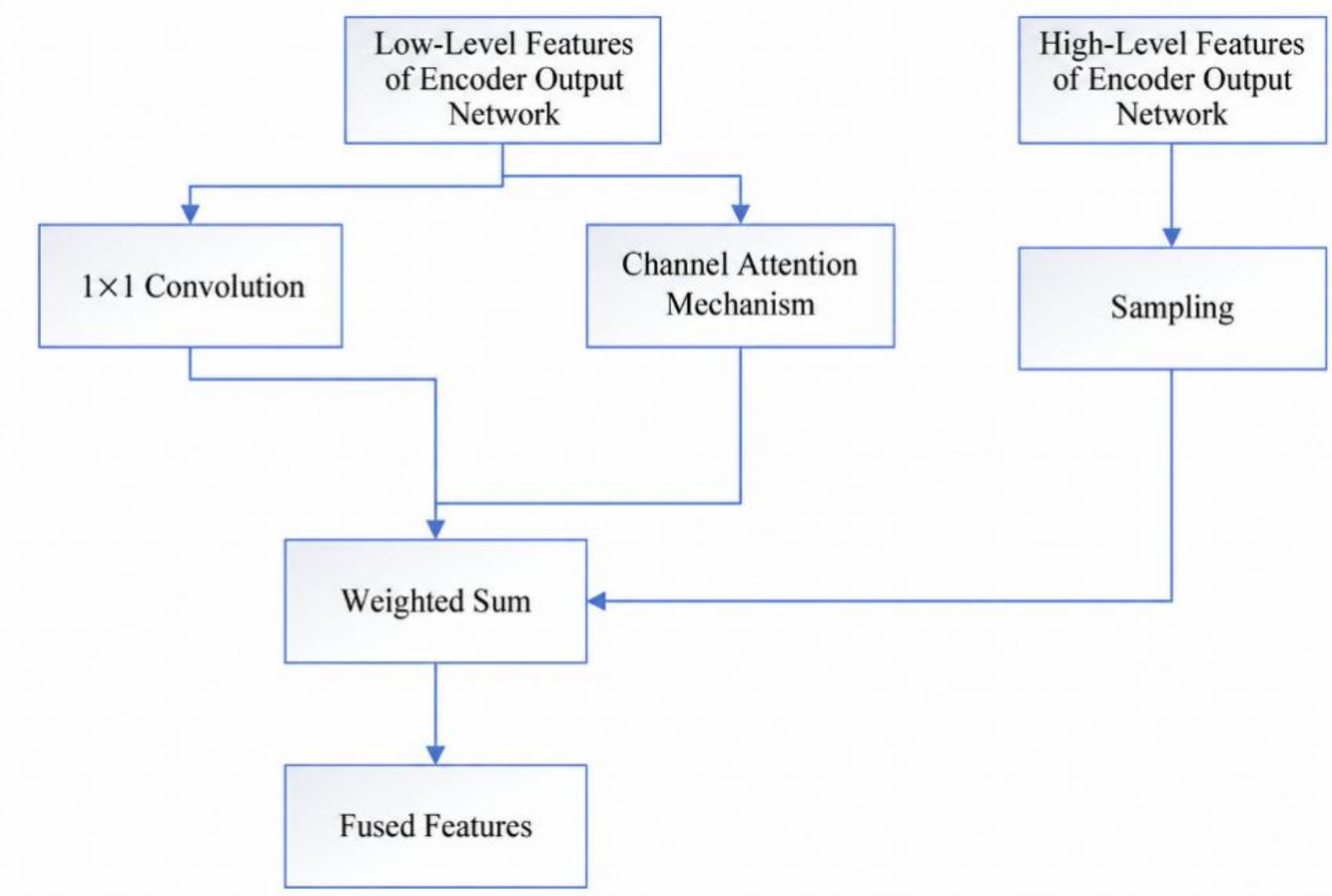

Figure 1 Workflow of Multi-scale Feature Fusion Module

Dynamic Blur Kernel Estimation Module

(1) Motion Trajectory Prediction Network

Temporal Convolutional Network (TCN) or optical flow networks such as PWC-Net are adopted to model the blur formation process and output pixel-wise motion vector fields.

Input: Blurred image sequences or temporal correlation assumptions of single images.

Output: 6-DOF camera motion parameters or non-rigid motion fields.

(2) Physics-Constrained Loss Function

Kernel consistency loss is introduced to guarantee the physical correlation between the predicted motion field and blurred images, as shown in Formula (1):

$$\mathcal{L}_{phy} = \| B - I \otimes K \|_2^2 \quad (1)$$

where $B$ denotes the blurred image, $I$ represents the sharp image, and $K$ stands for the predicted blur kernel.

Attention Enhancement Module

(1) Spatial-channel Dual AttentionChannel attention dynamically adjusts the weights of feature channels via Squeeze-and-Excitation (SE) Block.Spatial attention focuses on blurred boundary regions by adopting deformable convolution.

(2) Frequency-domain Guided Attention Fast Fourier Transform (FFT) is performed on feature maps, and band-pass filters are designed in the frequency domain to enhance the reconstruction of high-frequency edge components.

### 3.2 Optimization of Visual Large Model Algorithms and Inference Acceleration in Computing Processes

Large vision models represented by Vision Transformer (ViT) serve as core components in the computational phase of perception-computation-control services, undertaking critical tasks including complex visual feature extraction and high-level semantic understanding[37]. Their inference performance directly determines the decision-making efficiency and control accuracy of the whole system. Nevertheless, the global attention mechanism and deep network structure of ViT bring high computational complexity and large memory consumption, resulting in excessively high inference latency on traditional CPUs[38], which fails to meet the real-time requirements of perception-computation-control applications.Taking vit_base_patch16_224 as the optimization target, this study adopts the MagicMind inference framework on the Phytium-Cambricon MLU370 heterogeneous computing platform[39]. Efficient acceleration of large vision models is realized through structural optimization, quantization adaptation and operator fusion. While maintaining inference accuracy, the proposed methods significantly reduce inference latency and provide core support for low-latency operation of perception-computation-control services. This section elaborates the research content from three aspects: optimization scheme design, experimental verification and result analysis.

#### 3.2.1 Lightweight optimization of model structure

The vit_base_patch16_224 model consists of 12 Transformer encoder layers, each containing 12 attention heads[40]. The input image is split into patches to generate a sequence of 196 image patches, and the time complexity of global attention computation is $O(N^2D)$.On the premise of ensuring the inference accuracy loss $\leq 3\%$, this study carries out lightweight transformation on the model as follows:

(1) Attention Head Sparsification: Based on the importance analysis of attention weights, the 12 attention heads in each Transformer encoder layer are screened, and 8 core attention heads with the greatest contribution to feature extraction are retained, reducing the attention computation by 33%.

(2) Embedding Dimension Compression: The image patch embedding vectors are

dimensionally reduced via 1×1 convolution, compressing the embedding dimension from 768 to 512, which significantly decreases the computational complexity of matrix multiplication.

(3) Structural Module Fusion[41]: The combined modules of layer normalization, multi-head attention, and residual connection in the Transformer encoder are structurally fused to reduce the read/write overhead of intermediate feature maps and improve computational parallelism.

### 3.2.2 MagicMind Framework Adaptation and Quantization Optimization

This section utilizes the MagicMind 1.0.1 inference framework to complete the format conversion, quantization optimization, and hardware adaptation of the ViT model. The specific process is as follows:

(1) Model Format Conversion: Export the trained vit_base_patch16_224.pt model to the ONNX format, remove the Dropout layers and gradient calculation modules used during the training phase, and retain the core inference computation graph. Parse the ONNX model via MagicMind's onnx_parser tool, and then convert it into a .mm file executable on the MLU370 using the mm_build tool. This maps the model to the MagicMind computation graph, supporting the recognition and adaptation of core operators such as convolution, matrix multiplication, and layer normalization.

(2) INT8 Quantization Optimization: Adopt the Post-training Quantization (PTQ) scheme, select 100 images from the ImageNet validation set as calibration data, and quantize the model weights and activations from FP32 to INT8. For precision-sensitive modules such as attention calculation, mixed-precision quantization (INT8+FP16) is employed to balance inference speed and accuracy[42].

## 3.3 BANG C Core Operator Development and Optimization

Based on the MagicMind 1.0.1 inference framework, this section completes the format conversion, quantization optimization, and hardware adaptation of the ViT model. The specific process is as follows:

(1) Model Format Conversion: Export the trained vit_base_patch16_224.pt model to the ONNX format[43], remove the Dropout layers and gradient calculation modules used in the training phase, and retain the core inference computation graph. Parse the ONNX model via the MagicMind onnx_parser tool, and then convert it into a .mm file executable on MLU370 using the mm_build tool. The model is mapped to the MagicMind computation graph, supporting the recognition and adaptation of core operators such as convolution, matrix multiplication, and layer normalization[44].

(2) INT8 Quantization Optimization: Adopt the Post-training Quantization (PTQ) scheme,

select 100 images from the ImageNet validation set as calibration data, and quantize the model weights and activations from FP32 to INT8. For accuracy-sensitive modules such as attention calculation, mixed-precision quantization (INT8+FP16) is employed to balance inference speed and accuracy.

# 4 Implementation plan

## 4.1 DeepDeblur Model Porting and MLU Platform Optimization

In the PyTorch environment, export the pre-trained model of the DeepDeblur algorithm as an intermediate TorchScript file. Fully preserve the model's network structure, parameters, and computational logic, preparing it for subsequent processing in the MagicMind inference framework. As shown in Figure 2, the flowchart for generating a MagicMind model starts with MagicMind parsing the PyTorch model into a network, then using the builder to optimize and generate the model, followed by choosing to serialize and save it as a model file for future use. After setting the data type, the parse method of Parse needs to be called to import the original model from the open-source framework, such as the PyTorch .pt model[45].

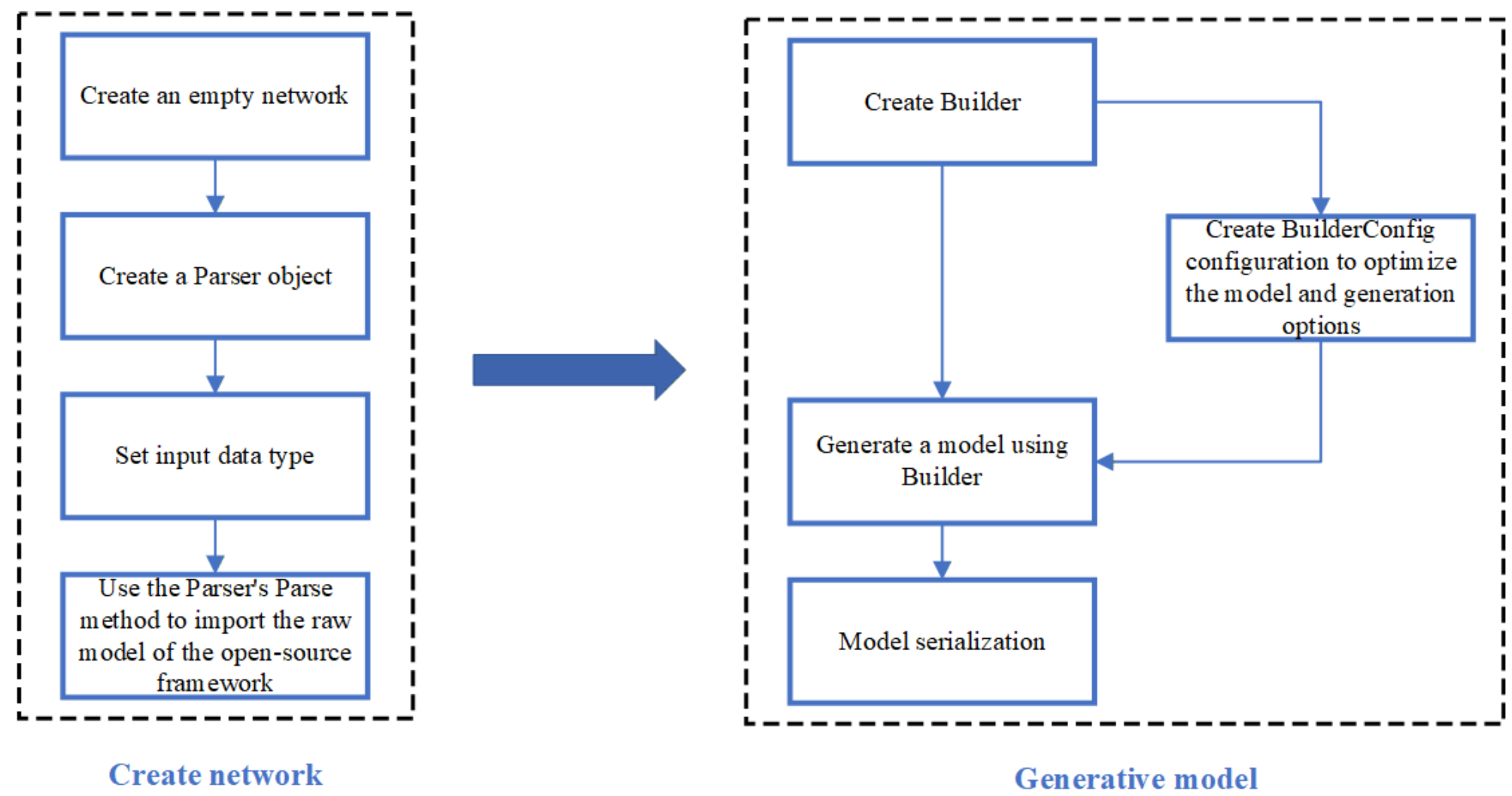

Figure 2 Generation Flow of MagicMind Model

The TorchScript model is converted into an inference-ready MagicMind model using the MagicMind 1.0.1 inference framework. During the conversion, the following optimizations are performed based on the hardware characteristics of the MLU370-S4[46]:

(1) Model Quantization: By converting model parameters from FP32 to INT8 during quantization, the memory footprint of the model is significantly reduced, and the inference speed is also remarkably improved. Figure 3 shows the comparison of memory usage and inference time before and after model quantization.

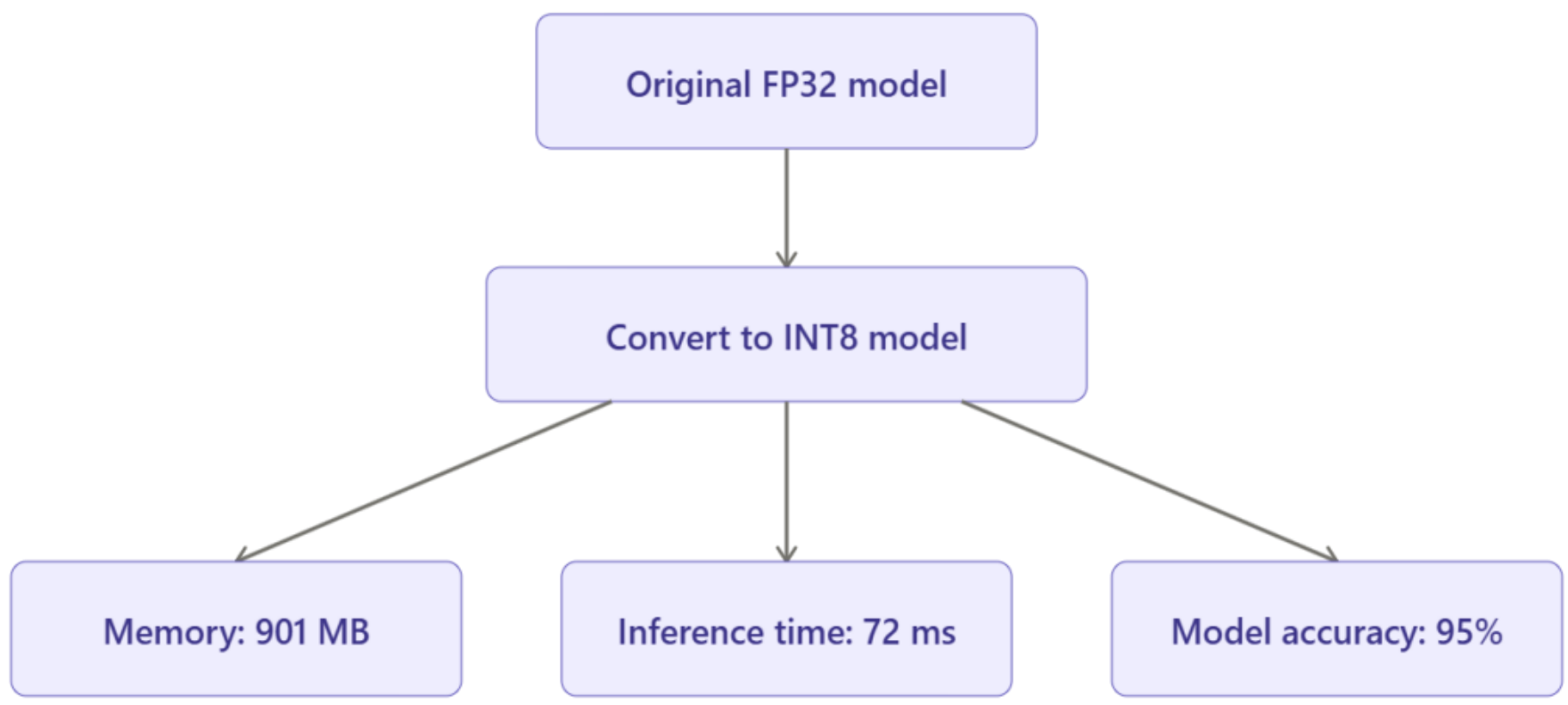

Figure 3 Comparison of memory usage and inference time after model quantization

(2) Operator Fusion and Memory Optimization: Consecutive convolution, activation, and normalization operators in the model are fused into a single composite operator to reduce the overhead of data migration between operators. The feature map memory layout is adjusted to the NHWC format to adapt to the access pattern of MLU vector computing units and improve memory bandwidth utilization.

After creating a builder, you can choose whether to create a BuilderConfig to configure model generation options. If no BuilderConfig is created, MagicMind generates the model using default parameters. If a BuilderConfig is created, you can manually configure parameters including device selection and model optimization.

The experimental flow is shown in Figure 4. First, images are randomly selected from the dataset as input and fed into the corresponding correction algorithm model. For deep learning-based algorithms, model loading and initialization are performed first; for traditional algorithms, processing follows their specific computational steps. Then, the corrected images output by the algorithm are obtained and compared with the corresponding ground-truth sharp images in the dataset to calculate various evaluation metrics. The above steps are repeated until all images in the test set are processed. Finally, statistical analysis is performed on all metrics to obtain the performance evaluation results of the algorithm.

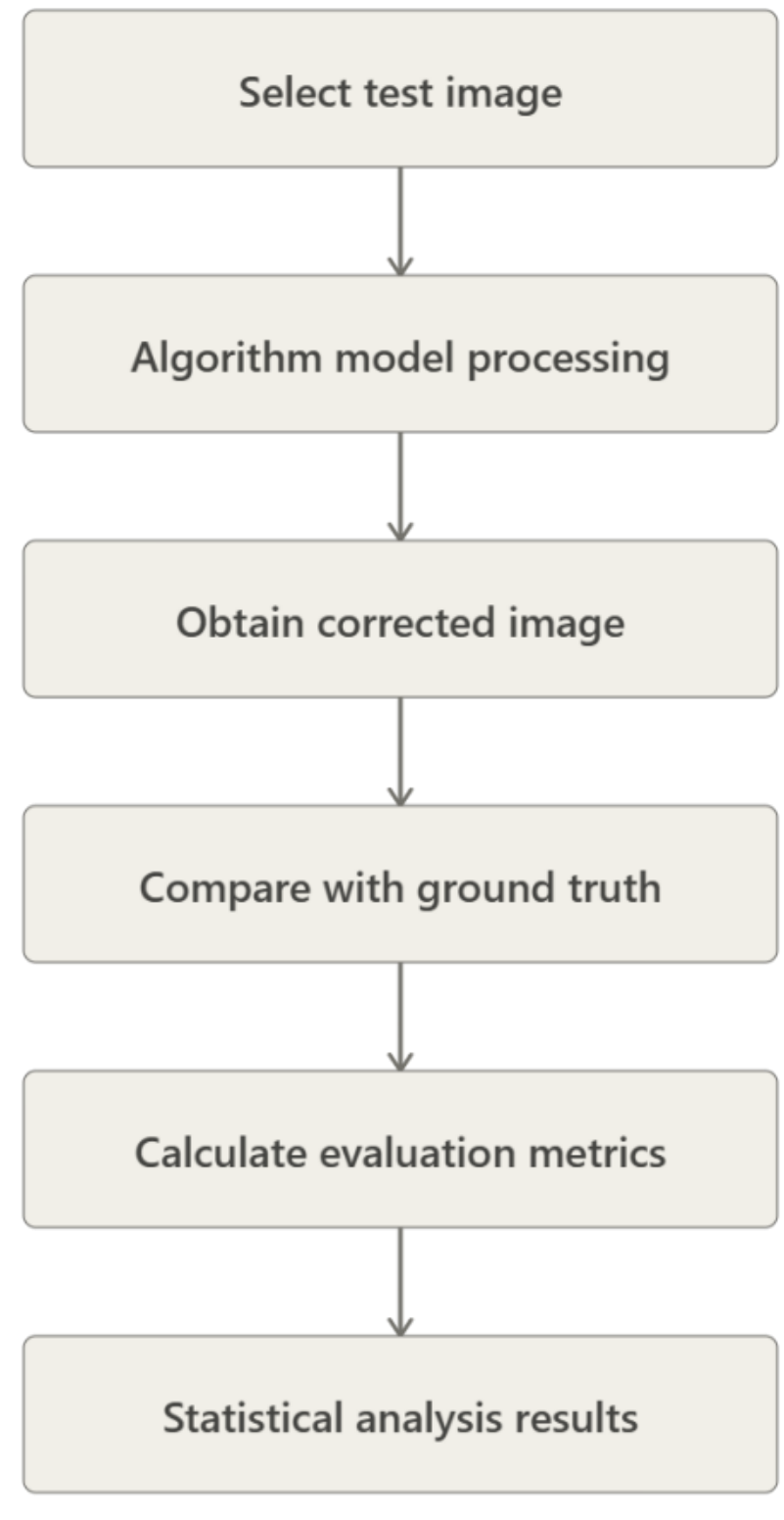

Figure 4 Experimental Flow Chart

## 4.2 Integrated Deployment and System-Level Scheduling Optimization of Single NPU Chip

Traditional perception-computation-control systems deploy perception, computation and control modules separately on different processors, resulting in frequent cross-device data transmission, which becomes the core bottleneck of system latency. To solve this problem, this study takes a single NPU chip as the core support and proposes an on-chip fusion architecture. All computing-intensive tasks throughout the perception-computation-control process are integrated and executed on a single MLU370 acceleration card, realizing on-chip data closed-loop and pipeline collaborative processing.

### 4.2.1 Single-card internal fusion architecture design

The single-card on-chip fusion architecture designed in this study adopts a four-layer hierarchical structure consisting of master scheduling layer, acceleration computing layer[47], data storage layer and interaction interface layer. The overall architecture is illustrated in Figure 5.

Master Scheduling Layer: Undertaken by Phytium FT-2000/4 CPU, it is responsible for receiving external sensor data, completing lightweight data verification, dynamically allocating MLU computing resources, issuing task instructions, and distributing final control commands to robot actuators to form closed-loop control.

Acceleration Computing Layer: Deployed on the MLU370 chip, it is functionally divided into three dedicated modules, namely preprocessing computing unit, model inference unit and control computing unit. They respectively execute image distortion correction, ViT model inference and inverse kinematics solution, serving as the core execution units for perception-computation-control services.

Data Storage Layer: Managed by partitioning the 24 GB video memory of MLU370 into input buffer area, intermediate result area and output buffer area. The intermediate result area supports cross-module data sharing among the three computing units. Preprocessed images and intermediate model feature maps can be directly transmitted on-chip without being transmitted back to CPU via PCIe interface, thus eliminating cross-device data transmission overhead.

Interaction Interface Layer: It realizes high-speed data interaction between CPU and MLU through PCIe 3.0 X16 interface, and achieves task synchronization via thread communication mechanism to ensure the timing consistency of pipeline execution.

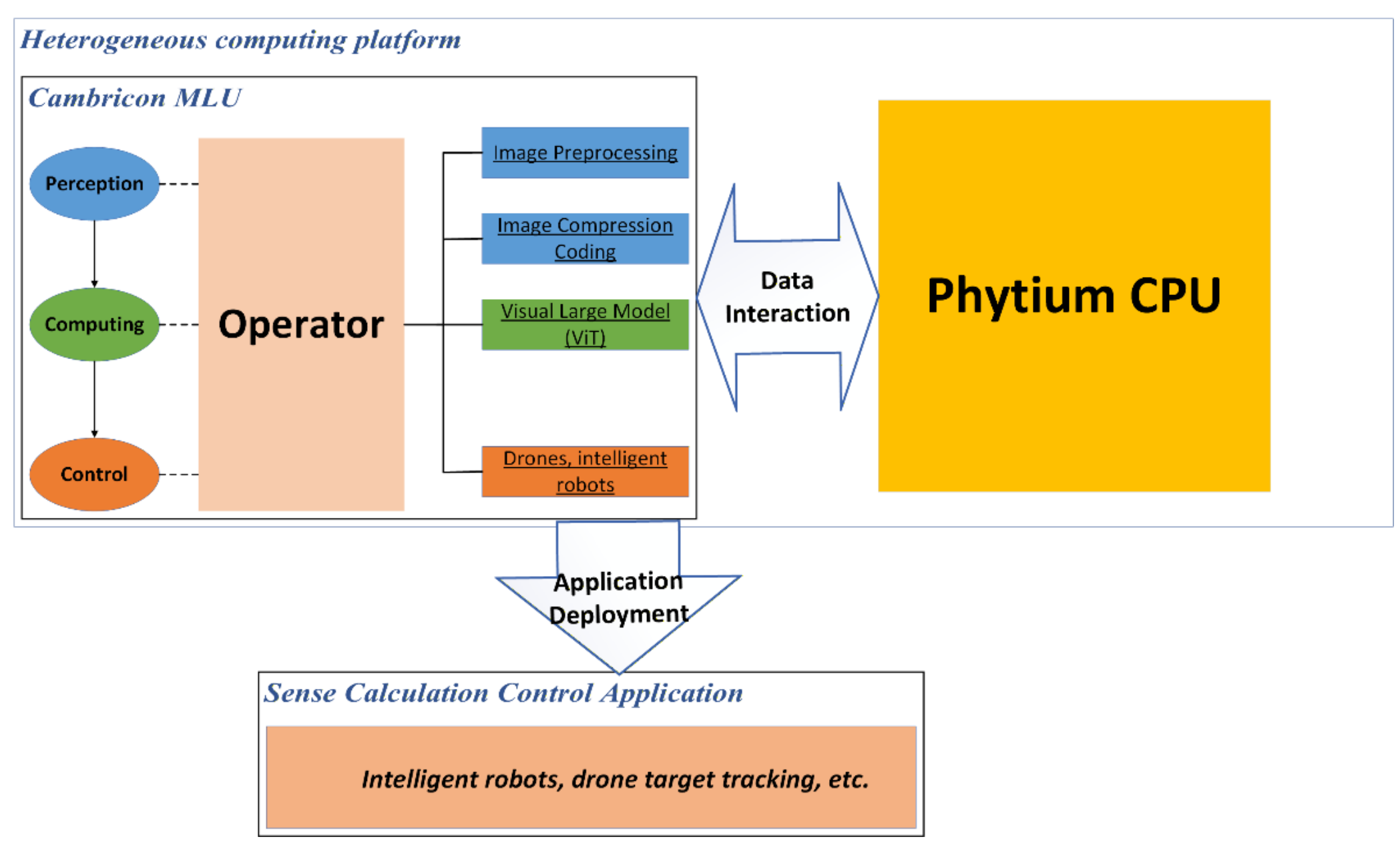

Figure 5 Integrated Deployment of Perception-Computation-Control Services on Heterogeneous Computing Platform

## 4.2 Full-process assembly line collaborative work mechanism

Based on the on-chip fusion architecture, this study designs a three-stage pipeline collaborative working mechanism of preprocessing-inference-control to realize full-process parallel processing[48].

Stage 1: Image data captured by the camera is transmitted to the MLU input buffer via the PCIe interface. The CPU issues preprocessing instructions, and the preprocessing computing unit starts distortion correction for the current frame.

Stage 2: The preprocessed images are stored in the intermediate result area. The model inference unit directly accesses on-chip data to launch ViT model inference, while the preprocessing unit processes the next frame in parallel.

Stage 3: Visual inference results are saved to the intermediate result area. The control computing unit reads the data to generate control commands synchronously. Meanwhile, the model inference unit continues the inference task of the previous frame, and the preprocessing unit processes the third frame.

With this pipeline mechanism, computing tasks of perception, computation and control are executed in parallel and overlapped inside the MLU chip. All data flows only within on-chip video memory without cross-device transmission, which greatly reduces end-to-end processing latency and PCIe interface bandwidth occupancy.

# 5 Introduction to the Experimental Environment

## 5.1 Hardware and software environment

### 5.1.1 Hardware Environment

(1) Cambricon MLU370 Edge AI Accelerator Card

The Cambricon MLU370 is a new-generation AI acceleration chip tailored for edge-side inference. Built on the MLUarch03 intelligent processor architecture, it is specially designed for perception-computation-control integrated services featuring high computing power, low latency and high concurrency. It is capable of supporting full-process computing-intensive tasks including image preprocessing, large vision model inference and control algorithm operators simultaneously.

As a typical edge-side version, the MLU370-S4 adopts a half-height and half-length single-slot design with controllable power consumption and strong scalability. It can work seamlessly with the Phytium FT-2000/4 processor via high-speed PCIe interface, meeting the deployment requirements of unmanned aerial vehicles, embodied intelligence and other scenarios. Figure 6 shows the MLU370-S4 edge AI accelerator card, and Table 1 lists its specifications.

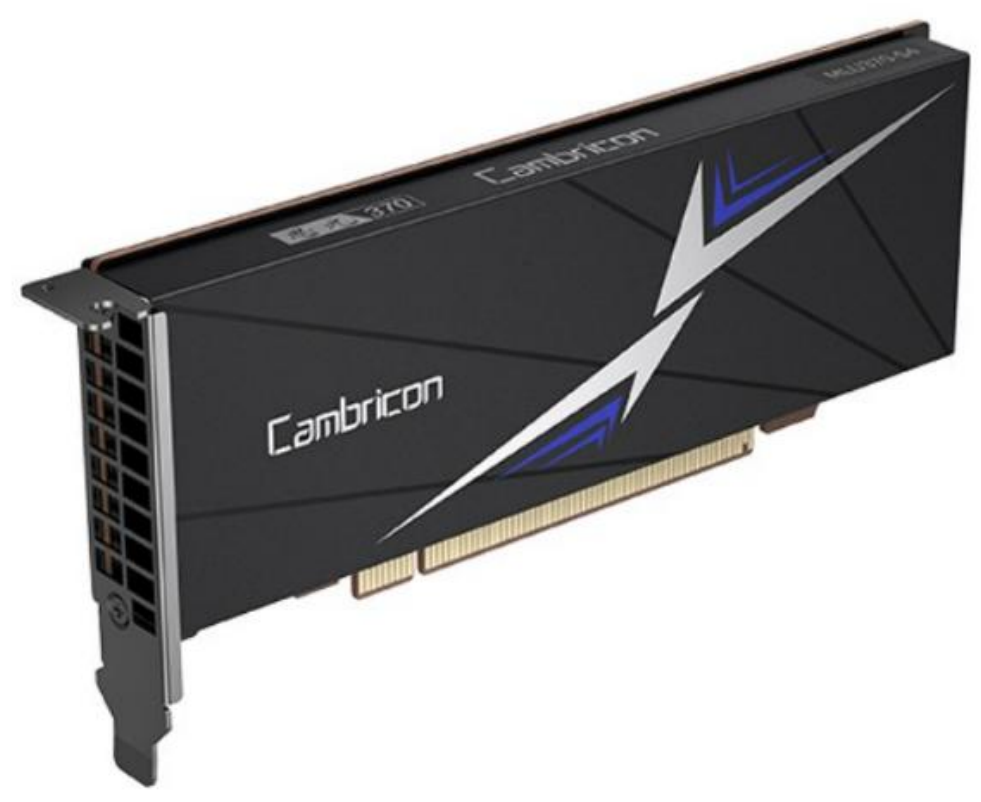

Figure 6 MLU370-S4 Edge Artificial Intelligence Accelerator Card

Table 1 Specifications of MLU370-S4

| | Parameter value |
|---|---|
| Peak performance | 192TOPS（INT8） |
| Memory Type | LPDDR5 |
| Memory capacity | 24GB |
| Image decoding | The maximum resolution supported for image encoding and decoding is 16384x16384 |
| Maximum thermal power consumption | 75W |

The MLU370 supports mixed-precision computing of INT8, FP16, BF16 and FP32, with a peak INT8 computing power of 192 TOPS. It can carry multiple workloads such as object detection, large vision model inference and control command generation simultaneously, satisfying the full-process computing power demands of embodied intelligent applications.

(3) Phytium FT-2000/4 Processor

Based on the ARM v8.2 instruction set architecture, the Phytium FT-2000/4 processor adopts a 4-core FT663 high-performance core design, as shown in Figure 7. Its hardware features and architectural design are well adapted to the requirements of low power consumption, high reliability and flexible expansion for perception-computation-control services.

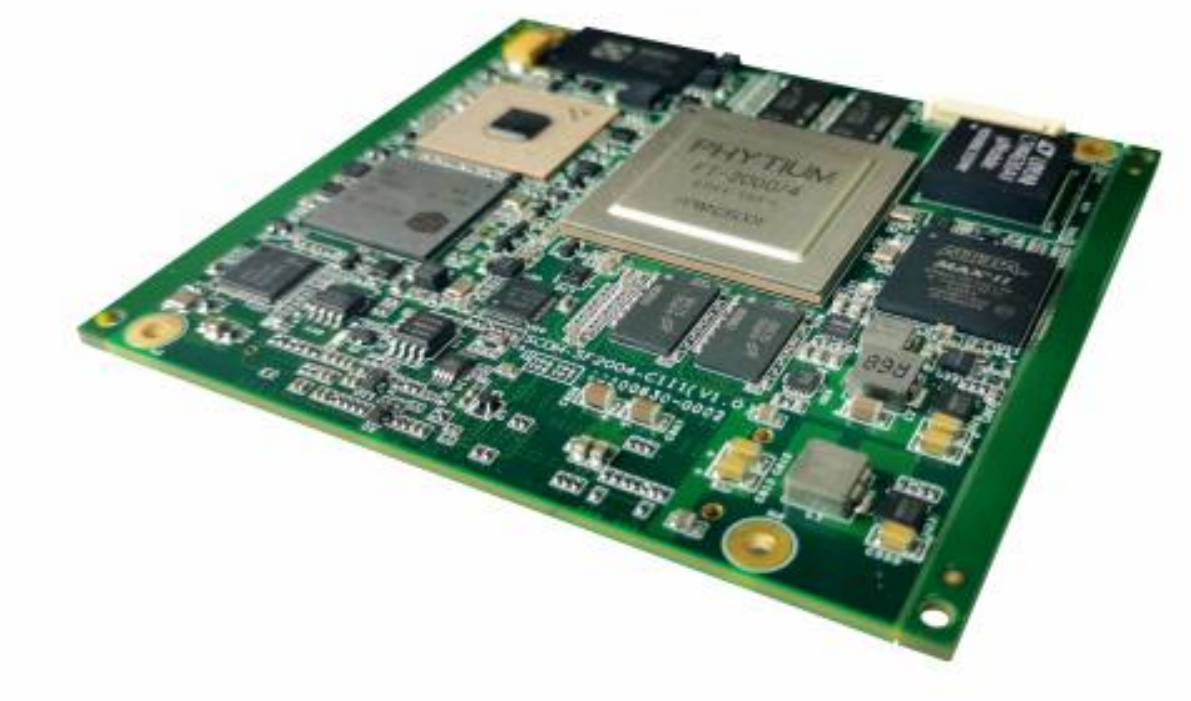

Figure 7 FT-2000/4 Processor

## 5.2 Test Dataset and Evaluation Metrics

In this study, comprehensive datasets for typical perception-computation-control application scenarios are selected, including:

500 frames of high-speed imaging motion-blurred distorted images (GOPRO dataset is shown in Figure 8);

Vision Transformer datasets ImageNet-21k and ILSVRC 2012 for large vision model experiments[49].

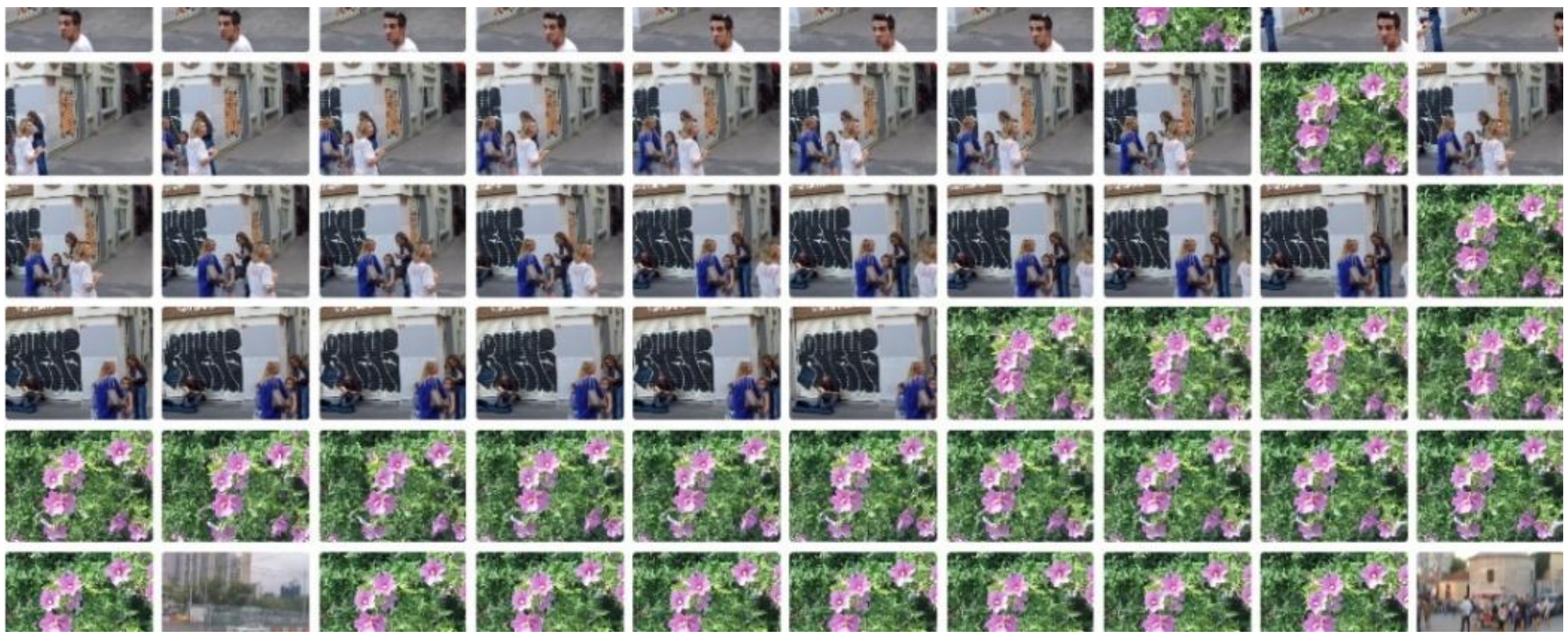

Figure 8 Image Examples of GOPRO Dataset

The experiments evaluate the overall performance from three indicators: end-to-end latency, processing accuracy and hardware utilization.

(1) End-to-end Latency: The total duration from image acquisition completion to control command generation, covering data preprocessing, model inference, control computation and data transmission time. The average single-frame latency is recorded for statistics.

(2) Processing Accuracy:

Image preprocessing accuracy: PSNR and SSIM for motion blur restoration;

Large vision model accuracy: Top-1 accuracy of ViT model;

Control accuracy: average error of robot joint angles and average deviation of target tracking.

(3) Hardware Utilization: Computing unit utilization rate and PCIe transmission bandwidth utilization rate.

# 6 Experimental Results and Analysis

## 6.1 High-speed imaging trailing distortion correction results

The generated DeepDeblur.mm model is adopted to perform distortion correction on high-speed motion-blurred images of running vehicles. The original images and corrected results are shown in Figure 9 and Figure 10 respectively.

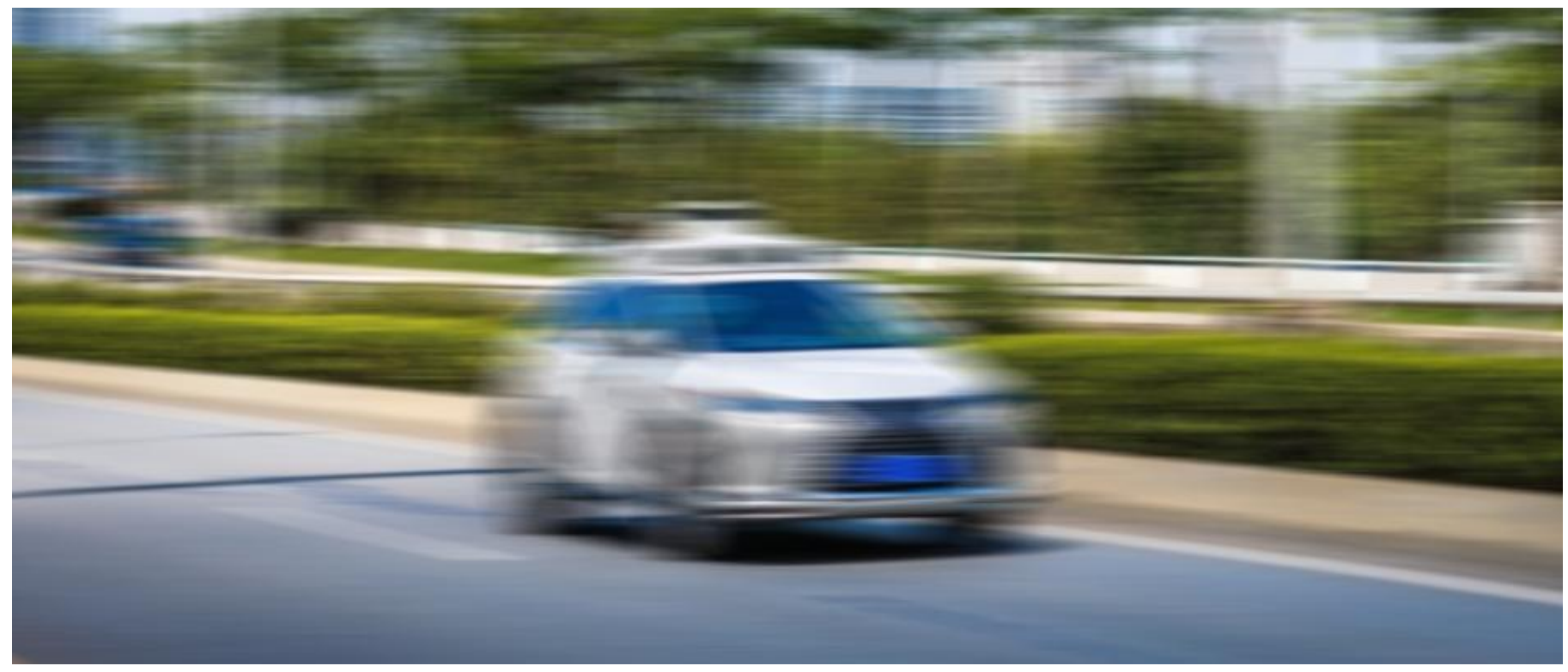

Figure 9 Original Car Image

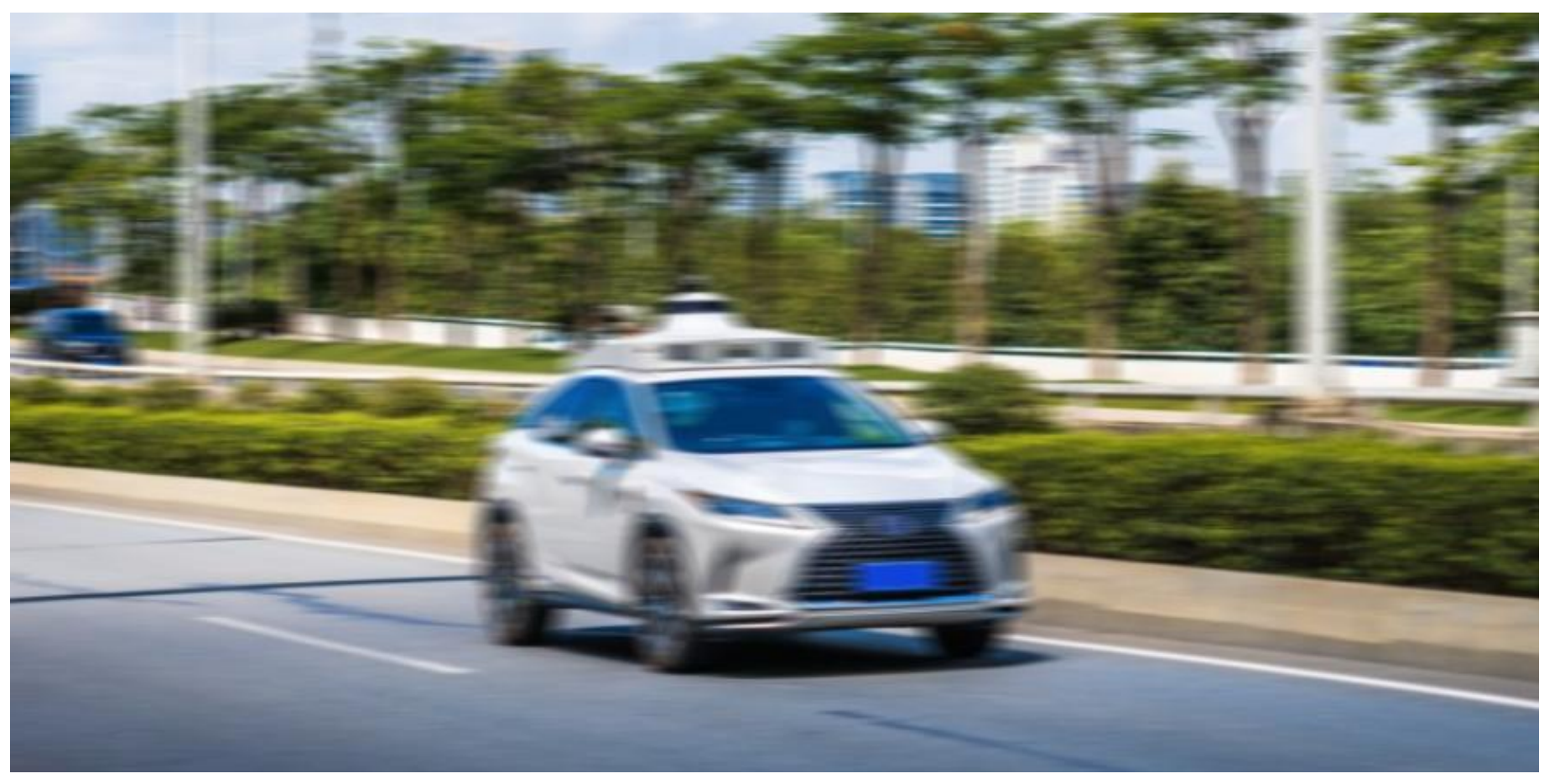

Figure 10 Vehicle Calibration Diagram

The comparison of single-frame image processing latency accelerated by MLU370 is shown in Figure 11. The single-frame processing latency of the original model is approximately 250 ms. After optimization via MagicMind and acceleration with the MLU370 accelerator card, the latency is reduced to around 70 ms, achieving an acceleration ratio of 3.65 times, which greatly cuts down inference latency.

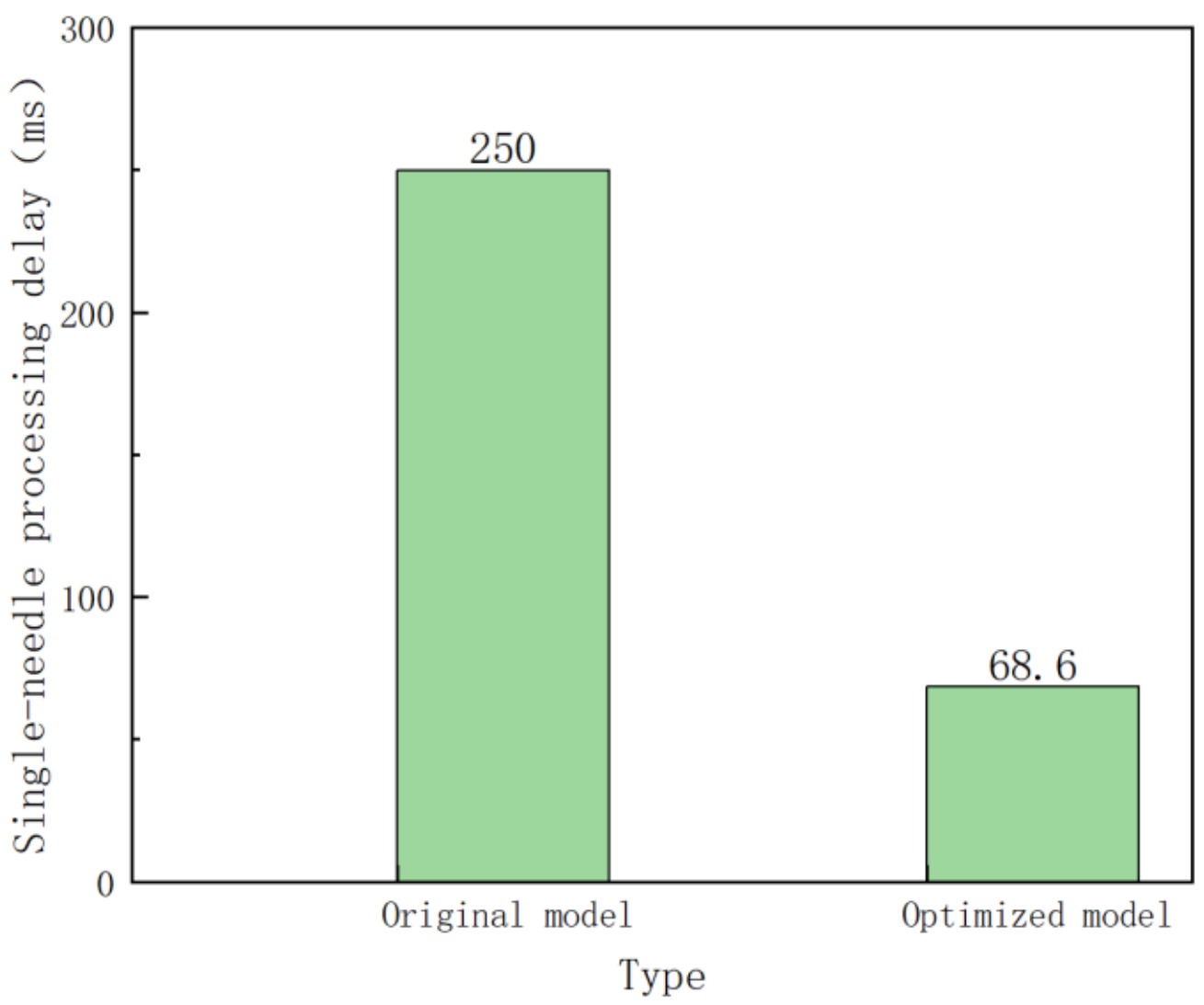

Figure 11 Model Single-Frame Image Processing Delay

## 6.2 Visual large model computation results

To verify the effectiveness of the ViT model optimization strategy, a heterogeneous experimental platform consisting of Phytium FT-2000/4 and Cambricon MLU370 is established. The ImageNet-1K validation set is used as test data to compare inference latency, recognition accuracy and hardware utilization of the model before and after optimization.

The single-frame processing latency of ViT model under pure CPU serial inference and MLU370 acceleration with MagicMind is shown in Figure 12.

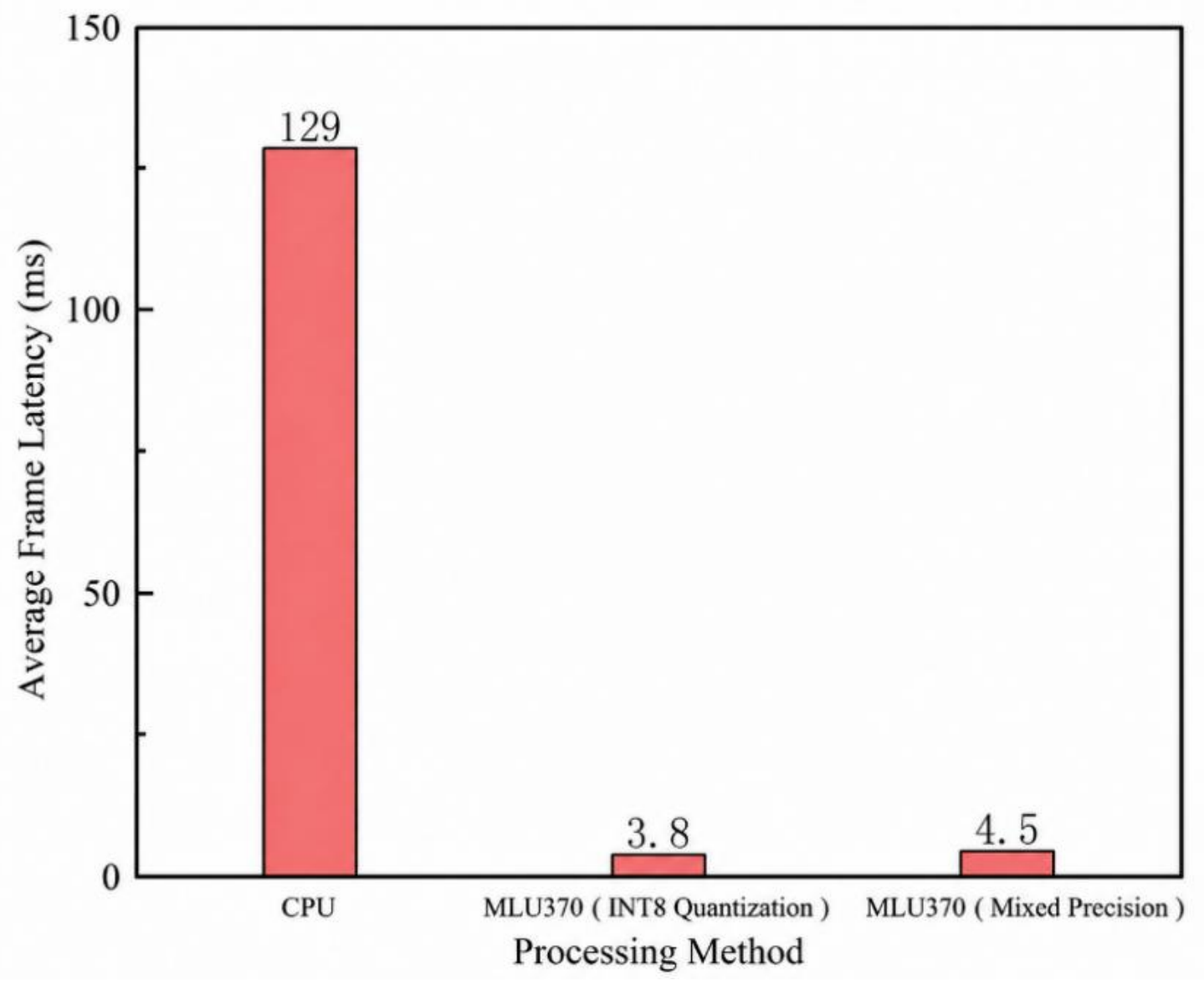

Figure 12 Comparison of Inference Latency of Visual Large Models

It can be seen from the experimental results that the average single-frame latency of pure CPU serial inference reaches 128.6 ms, while the acceleration scheme based on MLU370 greatly reduces the inference delay. The average single-frame latency of the INT8 quantized model drops to 3.8 ms with an acceleration ratio of 33 times, and that of the mixed-precision model is 4.5 ms with an acceleration ratio of 28 times, both fully meeting the real-time requirements of perception-computation-control services.

The inference accuracy of the ViT model before and after optimization is tested respectively, and the results are shown in Figure 13.

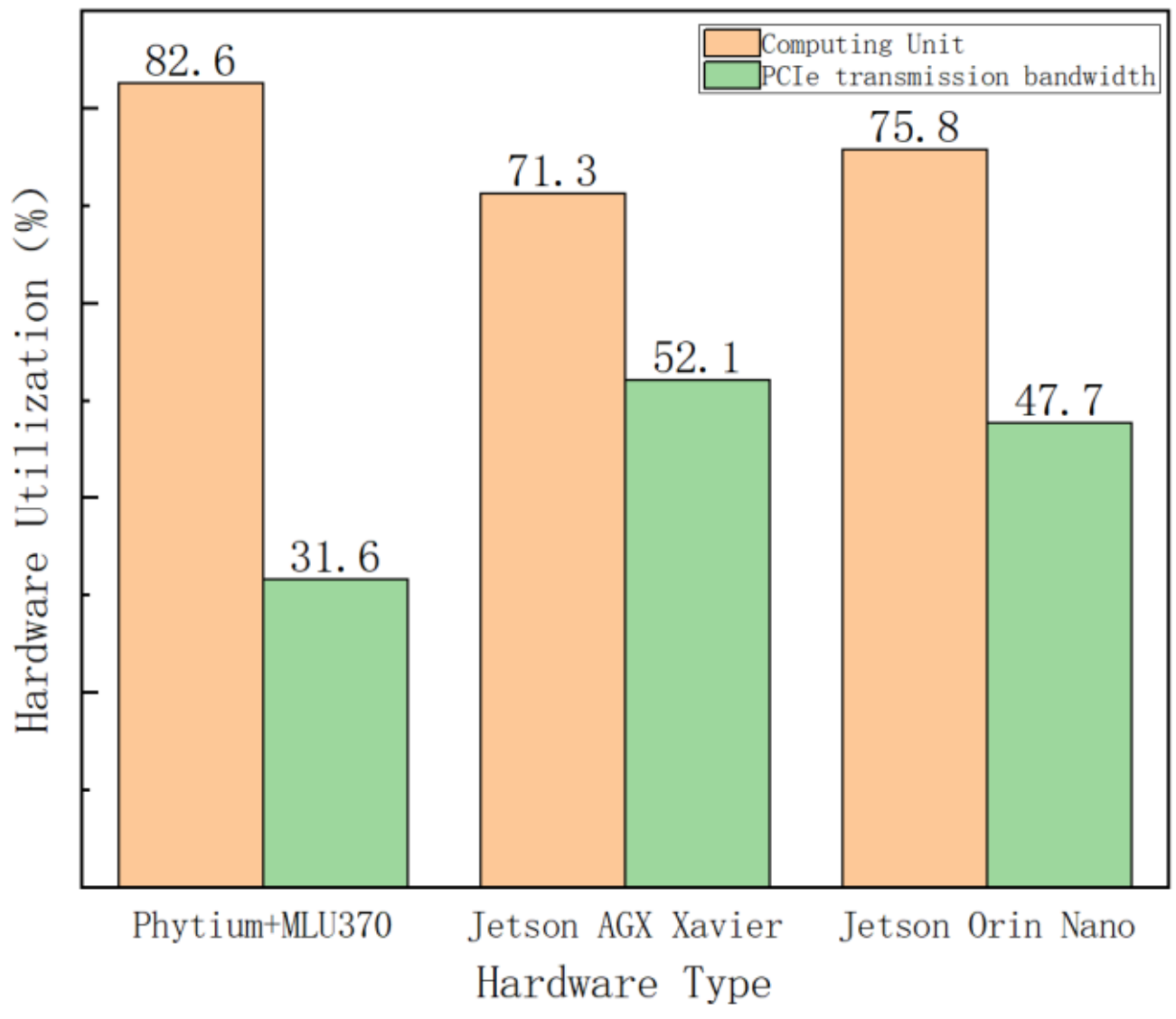

Figure 13 ViT Model Inference Accuracy

Experimental results show that the Top-1 accuracy of the INT8 quantized model is 75.8%, with an accuracy loss of 2.1% compared with the FP32 model running on CPU. The Top-1 accuracy of the mixed-precision model reaches 77.2%, with an accuracy loss of only 0.7%. Both accuracy losses are within an acceptable range.

## 6.3 Hundred Degree of Freedom Robot Control Results

Control algorithms serve as the core execution part of perception-computation-control services, which convert visual computing results into precise control commands for robots. The inverse kinematics solution for hundred-degree-of-freedom embodied intelligent robots requires solving high-dimensional nonlinear equations, involving dozens of matrix inversion and vector multiplication operations in a single solving step. Under traditional CPU serial computing mode, the time consumed for single-frame control command generation exceeds 120 ms, becoming a major bottleneck restricting system real-time performance.

To address this issue, this study develops native MLU operators for inverse kinematics solving based on the Cambricon BANG C heterogeneous programming model, and realizes hardware acceleration via parallel algorithm reconstruction.

The control latency before and after optimization is tested separately, and the experimental results are shown in Figure 14.

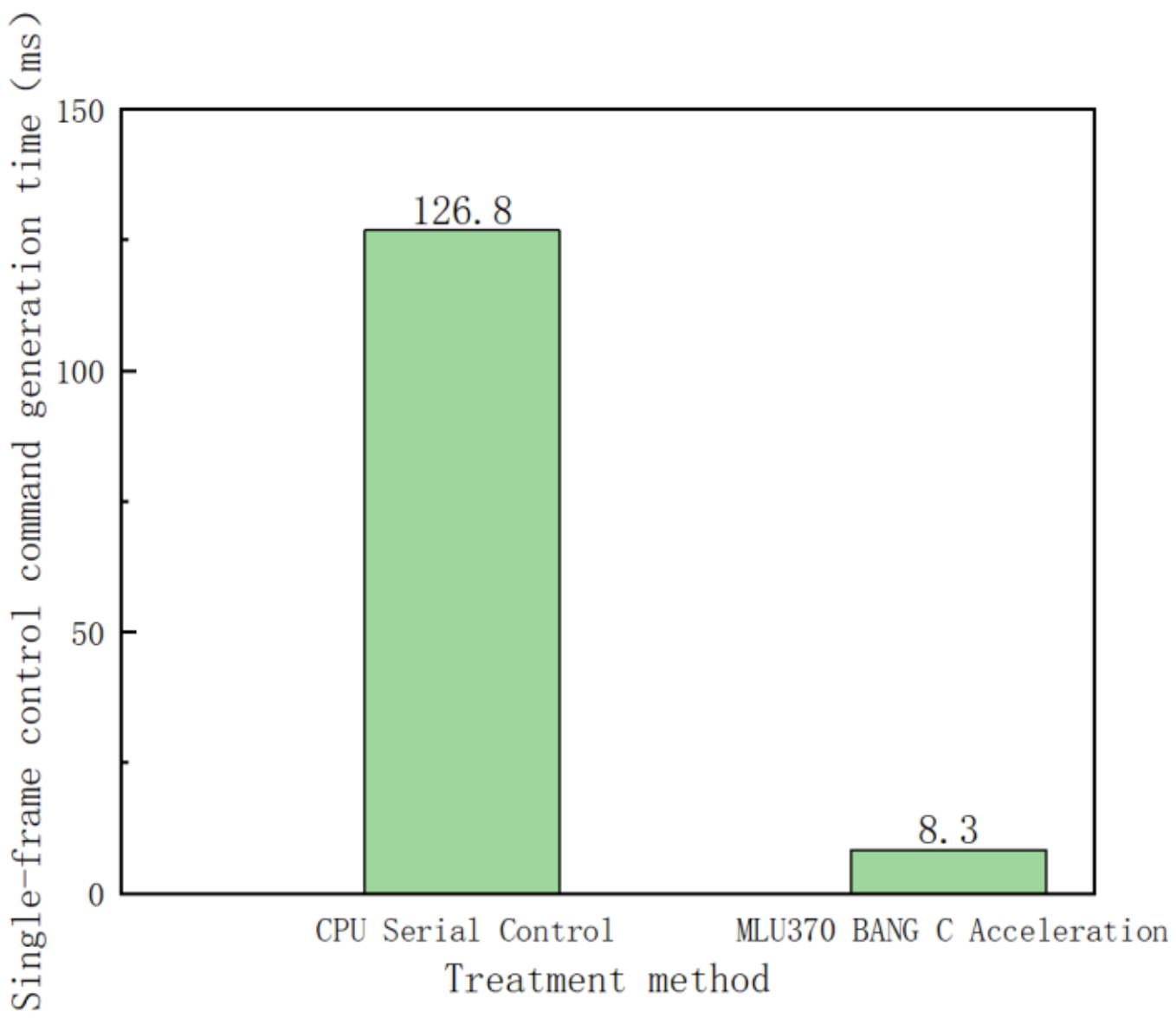

Figure 14 Comparison of Control Delay in a 100-DOF Robot

Experimental results indicate that the single-frame command generation time under CPU serial control reaches 126.8 ms. In contrast, the MLU370-based acceleration scheme drastically cuts down control latency, with the single-frame latency reduced to 8.3 ms under INT32 precision and an acceleration ratio of 15.28 times, fully satisfying the real-time operation requirements of perception-computation-control services.

The remarkable acceleration effect stems from three major optimized designs. Firstly, parallel reconstruction divides high-dimensional computing tasks into parallel subtasks to fully utilize the multi-thread parallel computing capability of MLU370. Secondly, native operators developed with BANG C are deeply optimized for hardware characteristics, effectively reducing instruction execution overhead. Thirdly, heterogeneous collaborative scheduling enables efficient cooperation between CPU and MLU, eliminating performance limitations of single hardware devices.

Taking robot joint angle error and target tracking deviation as evaluation criteria, the control accuracy of the MLU370 acceleration scheme suffers a slight decline yet remains within a reasonable range. Under FP32 precision, the joint angle error rises by 0.04 degrees and the target tracking deviation increases by 0.3 millimeters. Such accuracy loss is mainly caused by approximation errors introduced by fixed-point arithmetic and operator quantization.

### 3.3.2 Full-process end-to-end comparison

The average single-frame end-to-end latency of perception-computation-control processes on three platforms, namely FT-2000/4 plus MLU370, NVIDIA Jetson AGX Xavier and NVIDIA Jetson Orin Nano, is shown in Figure 15.

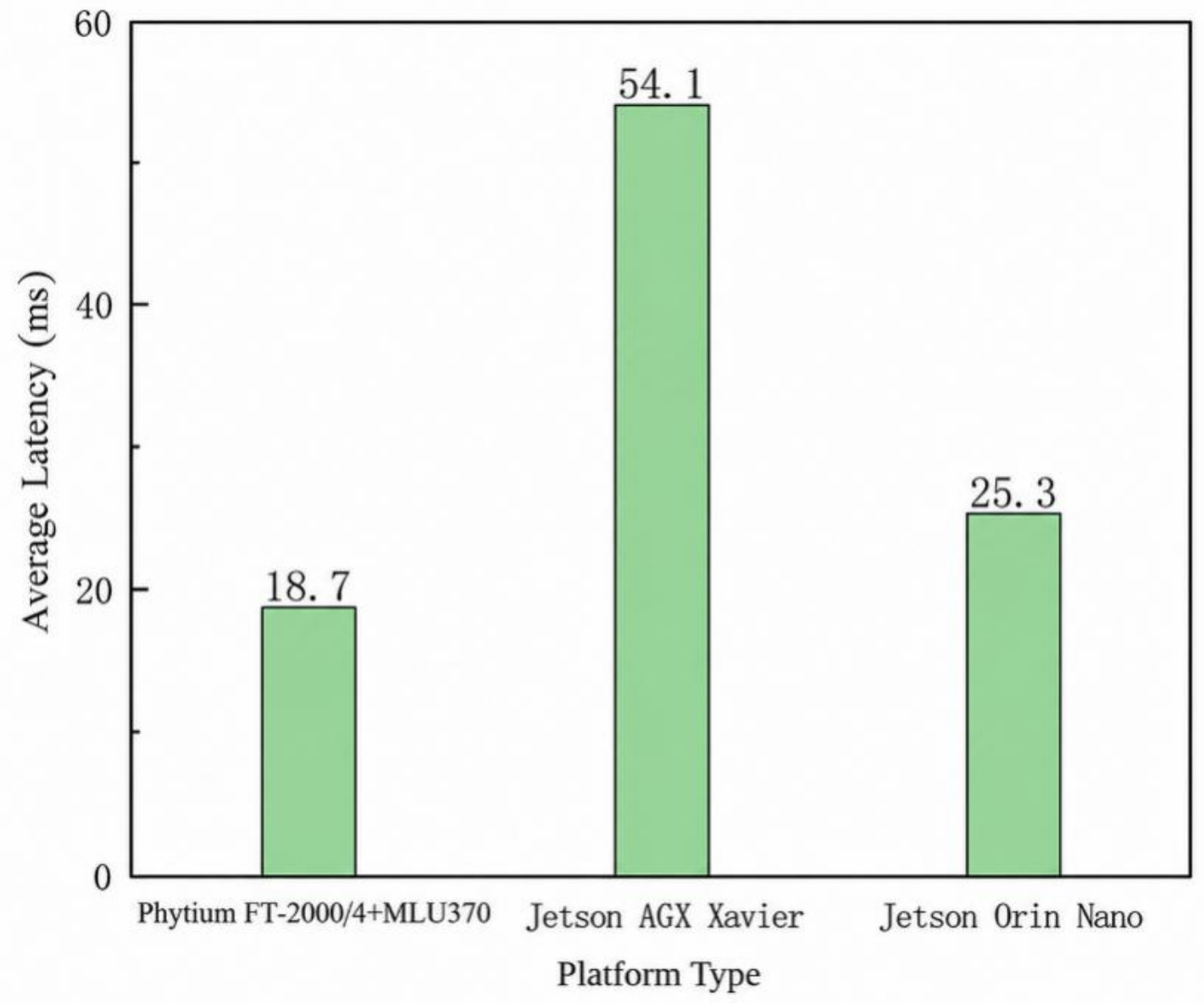

Figure 15 Single-frame average delay of the three major platforms

The acceleration ratios of each platform and the proportion of data transmission overhead (PCIe transmission time / total end-to-end time) are presented in Table 2.

Table 2 Comparison of Full-Process Latency of Sensory Computing Control

| Platform Type | Acceleration ratio | Proportion of data transmission overhead |
|---|---|---|
| Phytium FT-2000/4+MLU370 | 2.89 times | 0.3% |
| NVIDIA Jetson AGX Xavier | — | 18.6% |
| NVIDIA Jetson Orin Nano | 2.14 times | 12.4% |

It can be observed from the experimental results that the average single-frame latency of the Phytium+MLU370 platform is merely 18.7 ms, which is 2.89 times faster than Jetson AGX Xavier and 35.6% higher in efficiency than Jetson Orin Nano. Meanwhile, the data transmission overhead ratio of the MLU370 platform is only 0.3%[13], far lower than the two NVIDIA platforms, which fully verifies the advantages of the on-chip integration design.

The full-process accuracy test results of perception-computation-control services on the three platforms are listed in Table 3.

Table 3 Comparison of Full-Process Accuracy of Mental Arithmetic Control

| Accuracy indicator | Phytium FT-2000/4+MLU370 | Jetson AGX Xavier | Jetson Orin Nano |
|---|---|---|---|
| Trailing Correction PSNR（dB） | 27.87 | 28.12 | 27.63 |
| Trailing Correction SSIM | 0.876 | 0.883 | 0.871 |
| Top-1 Accuracy（%） | 75.8 | 76.2 | 75.5 |
| Average joint angle error（°） | 0.36 | 0.38 | 0.39 |
| Average deviation of target tracking（mm） | 3.0 | 3.1 | 3.2 |

It can be concluded from the above table that the overall accuracy of the Phytium+MLU370 platform is basically consistent with that of NVIDIA platforms, with only a slight drop of 0.5% to 1.0% in ViT inference accuracy, which is mainly caused by the influence of INT8 quantization on partial operators. In terms of control algorithm accuracy, there is barely any difference between MLU370 and NVIDIA platforms, and the former shows certain advantages. It proves that native operators developed based on BANG C can guarantee high numerical calculation accuracy for complex control tasks, and single-card integrated deployment brings no extra accuracy loss.

As shown in Figure 16, when perception, computation and control tasks run concurrently, the average utilization rate of MLU370 computing units reaches 82.6%, obviously higher than that of the two NVIDIA platforms. This demonstrates that the proposed on-chip fusion architecture and pipeline scheduling strategy can effectively improve the concurrent utilization efficiency of hardware resources and avoid idle resources and resource waste caused by single-task operation. Meanwhile, the PCIe transmission bandwidth utilization rate is only 31.6%, much lower than the comparison platforms. This verifies that the on-chip data closed-loop mechanism greatly reduces data interaction between CPU and NPU, and eliminates bandwidth bottlenecks and latency overhead brought by cross-chip data transmission from the system level.

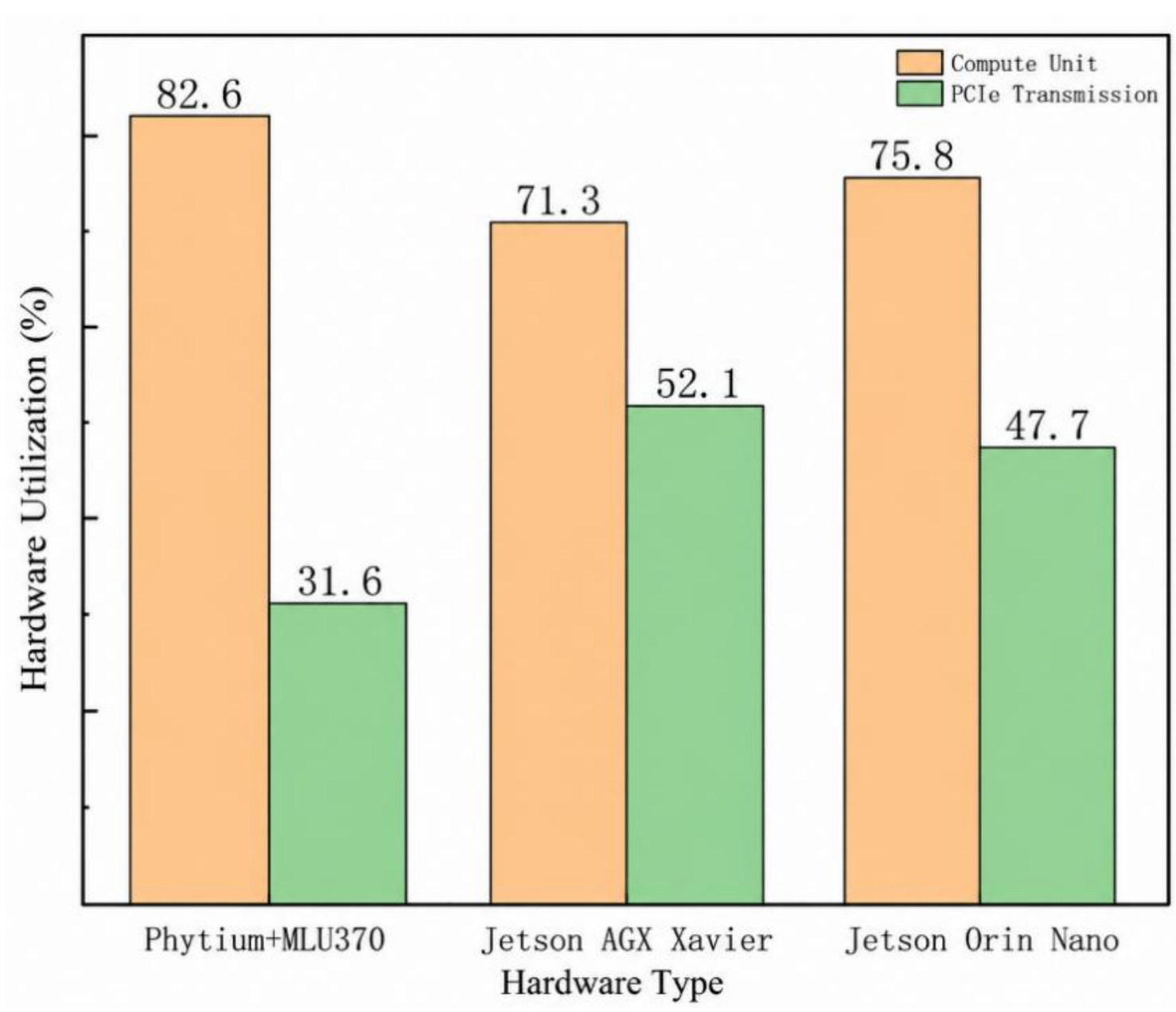

Figure 16 Comparison of Hardware Utilization Across Platforms

# 7 Conclusion

To achieve the domestic deployment of high-throughput and low-latency perception-computation-control services, this chapter adopts a single NPU chip as the core support. By carrying out algorithm optimization and hardware adaptation in perception, computation and control modules, the single-card integrated deployment of high real-time perception-computation-control service flows is realized. In the perception module, MLU hardware adaptation and MagicMind framework optimization are adopted to achieve a 3.65-fold acceleration for high-speed motion blur image restoration. In the computation module, structural lightweight design, INT8 quantization and operator fusion are applied to optimize the ViT large vision model, which achieves a 33-fold inference acceleration with the accuracy loss controlled within 2.1%. In the control module, customized operators are developed based on BANG C, realizing a 15.28-fold acceleration for the control calculation of hundred-degree-of-freedom robots and reducing the single-frame latency to 8.3 ms.

This paper constructs a single-card integrated architecture with on-chip data circulation and pipeline collaboration, integrating perception, computation and control functions into a single MLU370 accelerator card, and reducing the proportion of data transmission overhead to 0.3%. Experimental verification demonstrates that the end-to-end single-frame latency of the system is only 18.7 ms, with an acceleration ratio of 2.89 times compared with the NVIDIA Jetson AGX

Xavier platform. The utilization rate of MLU computing units reaches 82.6%, and its overall accuracy is equivalent to that of mainstream platforms. This scheme breaks the transmission bottleneck of traditional multi-device deployment, and provides a replicable technical solution for the localized engineering application of perception-computation-control services.